\pdfoutput=1 
\documentclass[aps,prl,showpacs,footinbib,twocolumn,final]{revtex4-1}
\usepackage{graphics,graphicx,color,ulem}
\usepackage[applemac]{inputenc}
\usepackage{amsmath,amssymb}

\newcommand{\smean}[1]{\langle #1 \rangle} 

\def\D{{\rm d}}                  

\def\ampAB{\delta{G}_\mathrm{AB}}

\begin{document}

\title{Ergodic {\it vs} diffusive decoherence in mesoscopic devices}

\author{Thibaut Capron$^{1}$, Christophe Texier$^{2,3}$, Gilles Montambaux$^{3}$, Dominique Mailly$^{4}$, Andreas D. Wieck$^{5}$ and Laurent Saminadayar$^{1,6,7,*}$}
\affiliation{$^{1}$Institut  N\'{e}el, B.P. 166, 38042 Grenoble Cedex 09, France}
\affiliation{$^2$Univ. Paris Sud ; CNRS ; LPTMS, UMR 8626 ; B\^at. 100, 
    Orsay F-91405, France}
\affiliation{$^3$Univ. Paris Sud ; CNRS ; LPS, UMR 8502 ; B\^at. 510, 
    Orsay F-91405, France}
 \affiliation{$^4$Laboratoire de Photonique et Nanostructures, route de Nozay, 91460 Marcoussis, France}
\affiliation{$^5$Lehrstuhl f\"{u}r Angewandte Festk\"{o}rperphysik, Ruhr-Universit\"{a}t, Universit\"atstra{\ss}e 150, 44780 Bochum, Germany}
\affiliation{$^6$Universit\'{e} Joseph Fourier, B.P. 53, 38041 Grenoble Cedex 09, France}
\affiliation{$^7$Institut Universitaire de France, 103 boulevard Saint-Michel, 75005 Paris, France}

\date{October 12, 2012}


\begin{abstract}
We report on the measurement of phase coherence length in a high mobility two-dimensional electron gas patterned in two different geometries, a wire and a ring. 
The phase coherence length is extracted both from the weak localization correction in long wires
and from the amplitude of the Aharonov-Bohm oscillations in a single ring, in a low temperature regime when decoherence is dominated by electronic interactions. We show that these two measurements lead to different phase coherence lengths, namely $L_{\Phi}^\mathrm{wire}\propto T^{-1/3}$ and 
 $L_{\Phi}^\mathrm{ring}\propto T^{-1/2}$. This difference reflects the fact that the electrons winding around the ring necessarily explore the whole sample (ergodic trajectories), while in a long wire the electrons lose their phase coherence before reaching the edges of the sample (diffusive regime).
\end{abstract}

\maketitle

Understanding the mechanisms of decoherence is a major issue in mesoscopic physics,  as it limits the existence of electronic interference effects~\cite{Imry,MontambauxBook}. The experimental determination of this quantity is thus of crucial importance. At relatively high temperature (typically above $1~K$), the electronic phase coherence is limited by electron-phonon collisions~\cite{ChaSch86}. 
Contrarily, at low temperatures, it is governed by electron-electron interactions: the measurement of the phase coherence time $\tau_{\Phi}(T)$ thus provides an important check of the theoretical models concerning these electronic interactions in metals~\cite{PieGouAntPotEstBir03,Saminadayar07} 
and a probe for studying the fundamental question of the interplay between interaction and disorder.

As it is well-known that in diffusive conductors the temperature dependence $\tau_{\Phi}(T)$ depends on space dimensionality $d$, it has long been assumed that it depends {\it only} on $d$. However, it turns out that the situation is richer in  quasi-one-dimensional  systems ($d=1$),  where decoherence results mostly from processus involving small energy exchange between electrons, therefore implying long length scales. Indeed it has been shown in a pionnering work~\cite{AAK82} that in this case,  the decoherence mechanism can be seen as the result of the interaction between one electron and the fluctuating electric field due to other electrons;
the random phase accumulated by an electron along its diffusive trajectory increases with time with a characteristic time $\tau_\Phi^\mathrm{wire}(T)\propto T^{-2/3}$. Implicitly this analysis is valid for infinite quasi-1D wire where diffusion is not limited in space. It has successfully explained many experiments on quasi-1D metallic or semiconducting wires.

In contrast, in a wire of finite length $L$, long diffusive trajectories are obviously limited at this length scale $L$, what must necessarily affect the decoherence mechanism. 
This so-called \textit{ergodic} regime is reached when the diffusion occurs over times longer than the Thouless time $\tau_D=L^2/D$, where $D$ is the diffusion coefficient. 
The ergodic regime has never been explored experimentally in diffusive wires. One reason is that a transport experiment implies connecting the device to external contacts, therefore it is hardly possible to investigate this ergodic regime in a mesoscopic wire.

An alternative and appropriate device  consists in a  ring in which the study of quantum oscillations in a magnetic field naturally provides a selection mechanism between diffusive and ergodic trajectories, because the harmonics of the oscillations necessarily involve winding (and therefore ergodic) trajectories.  
Several works~\cite{LM04,TexMon05b,TexDelMon09,Treiber09,TreTexYevDelLer11} have predicted that the temperature dependence of the phase coherence time obtained from magnetoconductance (MC) oscillations behaves as $\tau_\Phi^\mathrm{ring}(T)\propto T^{-1}$ in contrast with the result for an infinite wire.
The present work constitutes the first experimental evidence for such behavior
in a single ring.

More precisely, 
the goal of this paper is to provide, through  magnetotransport measurements,
a detailed comparison of the phase coherence time measured in two different geometries, namely an infinitely long wire and a ring. 
In a long wire,  interferences between time-reversed electronic trajectories give rise
to the  weak-localization (WL) correction to the {\it average} conductance. Its magnetic field dependence provides a measure  of the phase coherence length $L_\Phi^\mathrm{wire}=\sqrt{D\tau_\Phi^\mathrm{wire}}$~\cite{AltAro81,MontambauxBook}.
In a single ring, the magnetic field reveals the \textit{sample specific} interference pattern: it
modulates the phase difference between electron paths in each arm, leading to the so-called Aharonov-Bohm (AB) \cite{RefAB} periodic oscillations of the conductance~\cite{WasWeb86}.
The amplitude of the AB oscillations, averaged over an appropriate range of magnetic field, is controlled by the phase coherence length $L_\Phi^\mathrm{ring}=\sqrt{D\tau_\Phi^\mathrm{ring}}$. In this article we present for the first time a quantitative analysis of the temperature dependence of $L_\Phi^\mathrm{ring}$ extracted from AB oscillations of a single weakly disordered ring.
Our experimental analysis shows that the two phase coherence lengths $L_\Phi^\mathrm{ring}$  and $L_\Phi^\mathrm{wire}$ present two different temperature dependences, thus confirming the importance of the diffusive {\it vs} ergodic nature of the electronic trajectories~\cite{footnoteMeydi}.

Samples have been fabricated from high mobility GaAs/AlGaAs heterostructures. These systems are intrinsically clean of magnetic impurities~\cite{Saminadayar07} and thus provide a large phase coherence length only limited by electronic interactions at low temperature~\cite{YasuPRL09}. Using electron beam lithography on polymethyl-methacrylate resist and shallow etching, we have patterned mesoscopic rings and arrays of wires in the two-dimensional electron gas (2DEG). In addition, a Hall bar allows the determination of the characteristics of the 2DEG: the electron density is $n_e=1.56\cdot10^{11}\,cm^{-2}$, 
the mobility $\mu_e=3.1\cdot10^{5}\,cm^{2}/V\cdot s$ and the elastic mean free path  $\ell_e^\mathrm{(2DEG)}=2.13\,\mu m$, leading to a diffusion coefficient  $D^\mathrm{(2DEG)} = 1700\, cm^{2}/s$.

\begin{figure}[h]
\includegraphics[width=8cm]{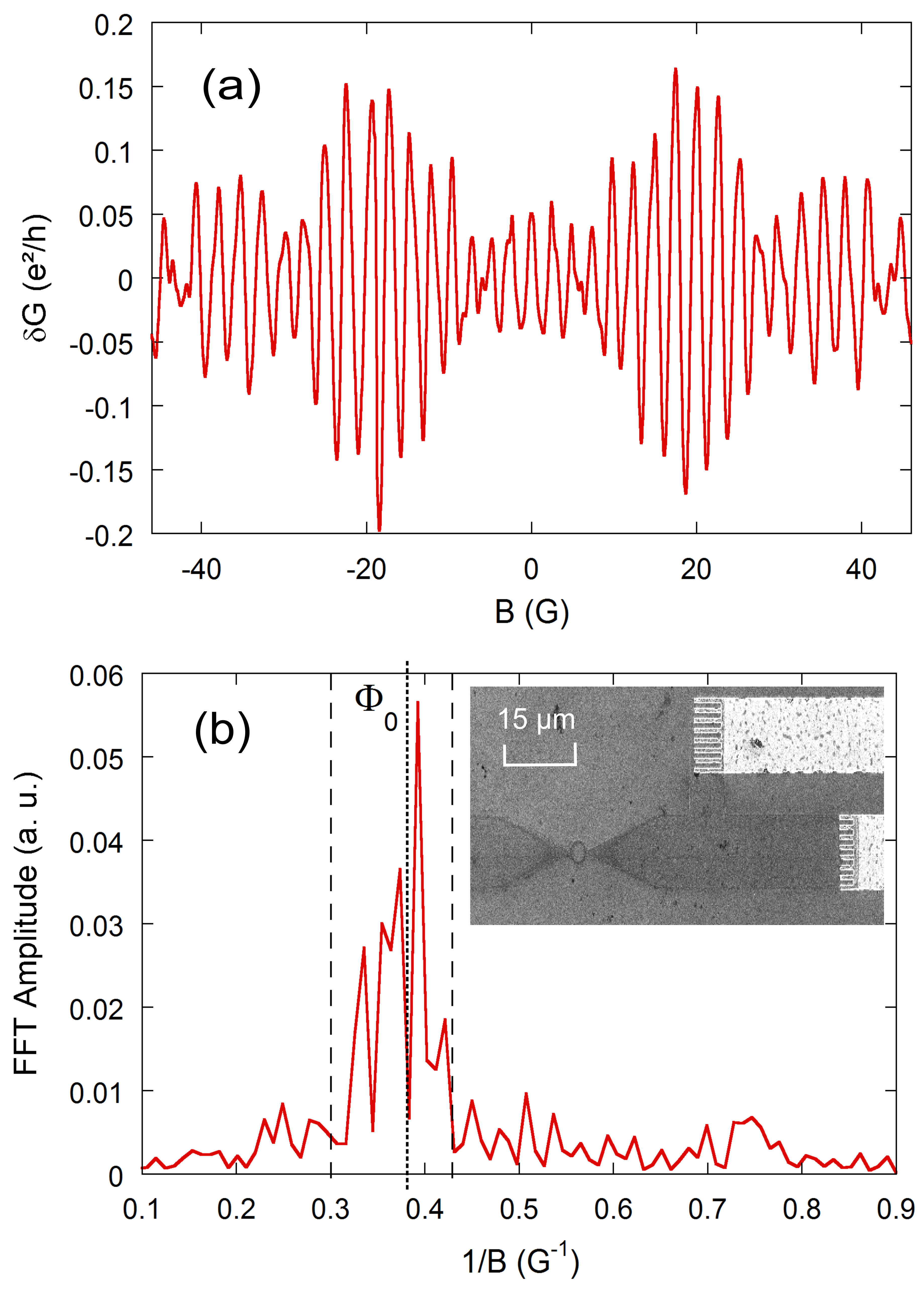}
\caption{(a): Magnetoconductance $\delta G$ of the ring measured at $55$~mK, after substraction of a smooth background signal. (b): Fourier Transform $\delta\widetilde{G}$ of the signal showing the $\Phi_{0}=h/e$ periodicity at $B_{0}$ (dotted line). The width of the main peak is represented by the dashed lines and corresponds to the values $1/B_{1,2}$ for the inner and outer $r_{1,2}$ radii of the ring. Inset: Scanning Electron Microscope image of the ring.}
\label{Fig:ring}
\end{figure}

Rings have a perimeter $L=13.6\,\mu m$, a lithographic width $w_\mathrm{litho}^\mathrm{ring}=740\,nm$, and are connected to large reservoirs, as presented in the inset of Fig.~\ref{Fig:ring}-b. Arrays of wires are composed of $20$ wires in parallel (in order to suppress conductance fluctuations) of width $w_\mathrm{litho}^\mathrm{wire}=800\,nm$ and length $L=150\,\mu m$. Note that due to depletion effects, the real widths $w^\mathrm{ring}$ and $w^\mathrm{wire}$ are somehow smaller~: 
it was shown in Ref.~\cite{YasuPRL09} for similar samples that depletion length is $\simeq190\:nm$ leading to $w^\mathrm{wire}\sim w^\mathrm{ring}\sim400\:nm$.
Analysis of the MC of the wires shows that the mean free path is close to the one of the 2DEG: $\ell_e\simeq \ell_e^\mathrm{(2DEG)}$.
In the two geometries, we thus probe the diffusive quasi~1D regime ($L_{\Phi}\gg \ell_e$, $w$).

MC measurements are performed on the ring and the wires simultaneously, using a standard $ac$ lock-in technique and a home made ultra low~noise amplifier ($0.4\,nV/\sqrt{Hz}$) at room temperature.

\begin{figure}[!h]
\includegraphics[width=8cm]{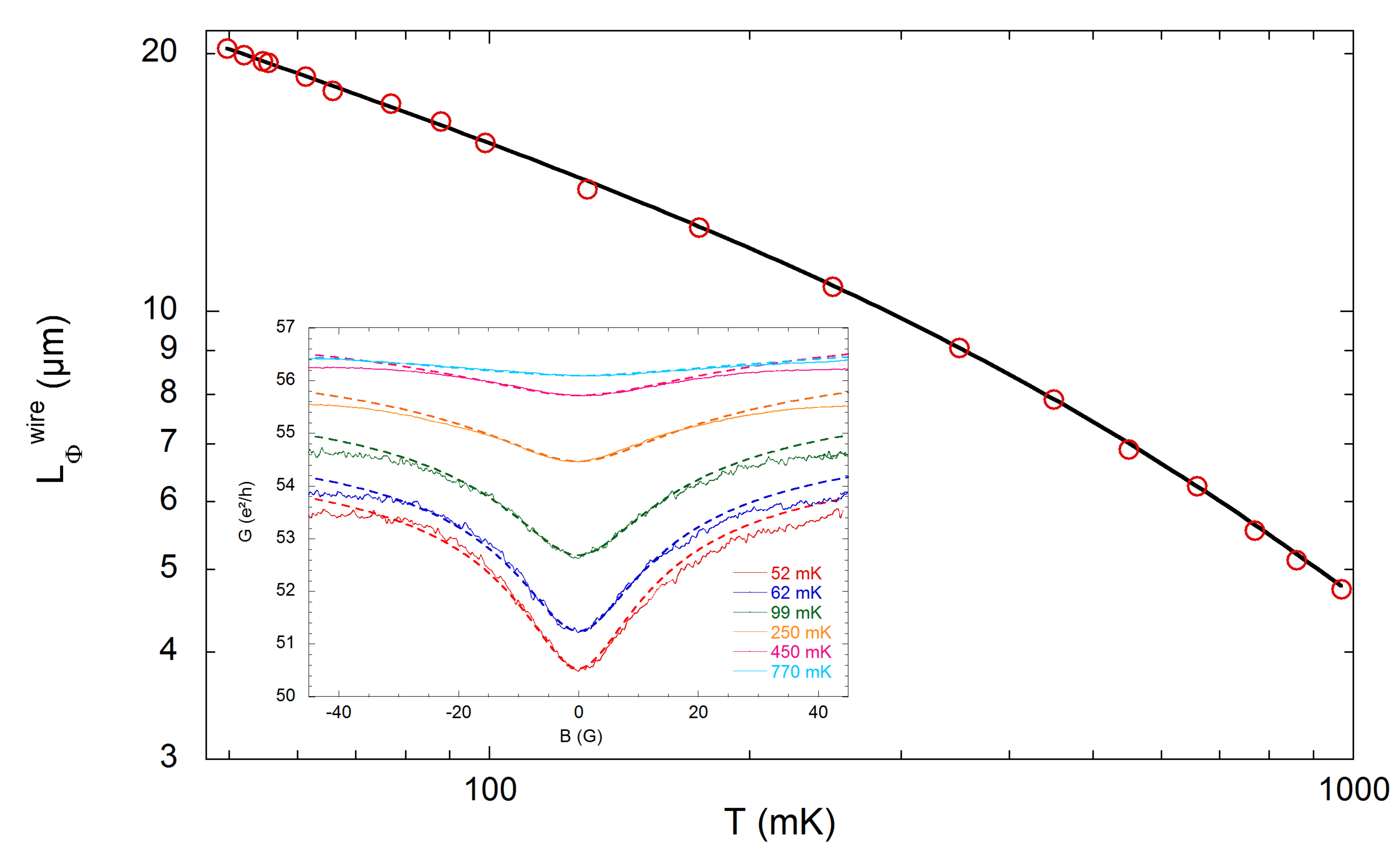}
\caption{Phase coherence length $L_{\Phi}^\mathrm{wire}$ as a function of temperature, extracted from weak-localization measurements. The solid line shows the theoretical fit to eq.~(\ref{eq:AAK}), using $\theta=2/3$. Inset: magnetoconductance of the wires measured at several temperatures. Dashed lines represent the fits to eq.~(\ref{eq:WL}).
}
\label{Fig:WL}
\end{figure}

We sweep the magnetic field perpendicularly to the sample on a span of $\pm 50\, G$ around zero-field. The conductance of the arrays of wires increases with magnetic field as expected, due to the suppression of the weak-localization correction. We then fit the MC with the 
well-known expression~\cite{AltAro81,MontambauxBook}: 
\begin{equation}
  \label{eq:WL}
  \Delta G = - \frac{2e^2}{h} \frac1L
  \left[ 
         \left(\frac{1}{L_\Phi^\mathrm{wire}}\right)^2 
      + \frac13\left(\frac{ew_*^\mathrm{wire}B}{\hbar}\right)^2 
  \right]^{-1/2}
  \:,
\end{equation}
where $w_*$ is an effective width accounting for flux cancellation effect  (for a wire~: $w_*=w$ if $\ell_e\ll w$ \cite{AltAro81} and 
$w_*\sim w\sqrt{w/\ell_e}$ if $\ell_e\gg w$ \cite{BeeHou88}).
As shown on the inset of Fig.~\ref{Fig:WL}, we can extract $L_{\Phi}^\mathrm{wire}$ at several temperatures ranging from $52\,mK$ to $1\,K$. The resulting temperature-dependence of $L_{\Phi}^\mathrm{wire}$ is displayed on Fig.~\ref{Fig:WL}. In this temperature range, decoherence is dominated by electronic interactions, and the phase coherence length is given for quasi~1D diffusive samples by~\cite{YasuPRL09}~:
\begin{equation}
  \label{eq:AAK}
    L_{\Phi} =\sqrt{ D\tau_\Phi }
   =\sqrt{\frac{D}{a\, T^{\theta}+b\, T^{2}}} 
\end{equation}
where, for a wire, $\theta= 2/3$
 and the factor $a$ is theoretically given by 
$D/a_\mathrm{theo}=2\big(m^*w^\mathrm{wire}D^2/\pi k_B\big)^{2/3}$
with $k_{B}$ the Boltzmann constant and $m^{*}$ the effective mass of the electrons, whereas the adjustable parameter $b$ takes into account 
interactions involving high energy transfers~\cite{footnoteZNA,AAK82,FA83,YasuPRL09}.
As shown on Fig.~\ref{Fig:WL}, $L_{\Phi}^\mathrm{wire}(T)$ is well described by Eq.~(\ref{eq:AAK}), setting $a_\mathrm{exp}$ as a fitting parameter.
A quantitative comparison of $a_\mathrm{exp}$ and $a_\mathrm{theo}$ for similar wires has been provided in Ref.~\cite{YasuPRL09} and has shown that the analysis of the WL leads to a reliable determination of $L_{\Phi}^\mathrm{wire}$~; we will not elaborate more on this point since the present analysis does not rely on it.

The magnetoconductance measured in the ring presents periodic oscillations. After substraction of a smooth background component, we obtain the characteristic Aharonov-Bohm oscillations shown on Fig.~\ref{Fig:ring}-a. These oscillations are $\Phi_{0}$-periodic with the magnetic flux, where $\Phi_{0} = h/e$. This corresponds to  a field periodicity $B_{0}=\Phi_{0}/(\pi r_{0}^{2})$ with $r_{0}$ the mean radius of the ring. A Fourier Transform of the signal is then performed using a rectangular window; this is presented on Fig.~\ref{Fig:ring}-b. A clear peak around $B_{0}=2.7\,G$ appears in the spectrum. From this value, we can extract $r_{0}=2.17\,\mu m$, in excellent agreement with the lithographic radius $r_{0}=2.13\,\mu m$.

In the following, we detail the procedure used to extract the amplitude of the AB oscillations. As the width of the arms of the ring is finite, all electron trajectories lying between the inner radius $r_{1}$ and the outer radius $r_{2}$ of the ring can participate to the oscillations. In addition, due to depletion inherent to the fabrication process (etching), the inner and outer radii are not the lithographic ones and the effective width of the arms of the ring $w^\mathrm{ring}$ is smaller than $w^\mathrm{ring}_\mathrm{litho}$. We observe a broadening of the Fourier peak between $1/B_{2}$ and $1/B_{1}$ (dashed lines in Fig.~\ref{Fig:ring}-b) corresponding to $B_{1}=\Phi_{0}/(\pi r_{1}^{2})$ and $B_{2}=\Phi_{0}/(\pi r_{2}^{2})$. From these results, we can extract the effective width $w^\mathrm{ring}=r_{2}-r_{1}=360\,nm$; it corresponds to a depletion length of $190\,nm$, in accordance with the value reported elsewhere using another method  on similar samples~\cite{YasuPRL09}. 
The conductance oscillations are Fourier transformed over a window $[-50\:G,+50\:G]$ (larger than the correlation field estimated to be $B_c\sim8\:G$). We clearly identify a peak corresponding to the AB $\Phi_0$-periodic oscillations (Fig.~\ref{Fig:ring}): 
in this case the AB amplitude is given by 
integrating this peak as 
$\ampAB^2\propto\int_{\mathrm{peak}\:h/e}\D K\,\delta\widetilde{G}(K)^2$ where $\delta\widetilde{G}(K)$ is the Fourier transform plotted in Fig.\ref{Fig:ring}(b). 
Theoretically, we expect that, for $L_{\Phi}^\mathrm{ring}\ll L$, $\ampAB$ depends on $L_{\Phi}^\mathrm{ring}$ as~\cite{WasWeb86}
\begin{eqnarray}
  \label{eq:HarmonicRing}
  \ampAB 
  \simeq 
  C\, \frac{e^{2}}{h}\, \frac{L_{T}}{L}\, 
  \left(\frac{L_{\Phi}^\mathrm{ring}}{L}\right)^{\eta}
  e^{ -L/2L_{\Phi}^\mathrm{ring} }
\end{eqnarray}
where $L_{T}=\sqrt{\hbar D/k_{B}T}$ is the thermal length, $L$ the perimeter of the ring and $C$ a constant of order $1$.
The exponent $\eta$ depends on the nature of the contacts. If the ring is isolated it is equal to $\eta=1/3$ \cite{TexMon05b,TexDelMon09,Tex07b}. We now justify that this value is appropriate in our samples. The  2D large contacts are connected to the ring through relatively narrow constrictions of width of the same order than $w^\mathrm{ring}$. Because the motion is ballistic at the scale of the contact (since $\ell_e>w^\mathrm{ring}$), forward scattering should be favored for an electron winding inside the ring, i.e. the probability to exit should be diminished, compared to the diffusive contact with $\ell_e\ll w^\mathrm{ring}$.
Note however that the value of $\eta$ will not affect strongly the fitting procedure for $L_{\Phi}^\mathrm{ring}$.

From the experimental data we can extract the phase coherence length in the ring $L_{\Phi}^\mathrm{ring}$ as a function of temperature. Note that Eq.~(\ref{eq:HarmonicRing}) is strictly valid only in the regime where $L_{\Phi}<\pi r_{0}$, the length of one arm of the ring. Indeed, 
in the opposite case $L_{\Phi}>\pi r_{0}$, 
electrons may explore the contacts with a high probability, what affects the decoherence and strongly modifies the $L_\Phi$-dependence of the AB amplitude~\cite{contacts,Tex07b,TexDelMon09}~:
for this reason, our analysis only holds above $T\simeq300\,mK$.

\begin{figure}[h]
\includegraphics[width=8cm]{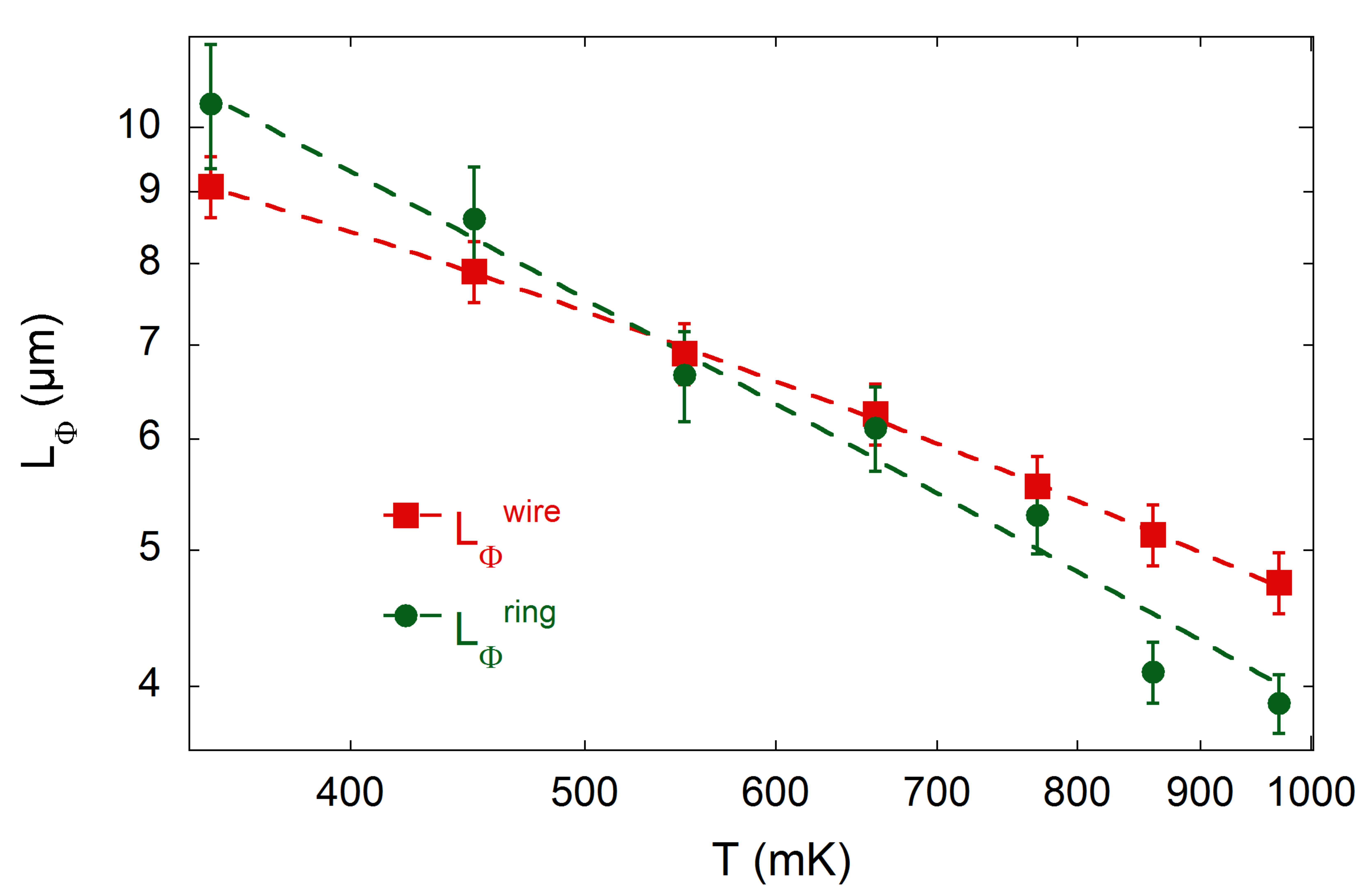}
\caption{Phase coherence lengths obtained in a wire, $L_{\Phi}^\mathrm{wire}$, and in a ring, $L_{\Phi}^\mathrm{ring}$, as a function of temperature. The dashed lines show the theoretical fit obtained with  eq. (\ref{eq:AAK}).
}
\label{LPhi}
\end{figure}

A direct comparison between $L_{\Phi}^\mathrm{wire}(T)$ and $L_{\Phi}^\mathrm{ring}(T)$ is presented on Fig.~\ref{LPhi}. We observe that the two phase coherence lengths differ both in {\it absolute value} and in {\it temperature dependence}. By fitting the data we obtain $L_{\Phi}^\mathrm{wire}\propto T^{-0.33\pm0.01}$ for the wire and $L_{\Phi}^\mathrm{ring}\propto T^{-0.49\pm0.09}$ for the ring \cite{footnoteLionel,HanKriPedSorLin01}. 
These exponents are in perfect agreement with the theoretical predictions 
$L_{\Phi}^\mathrm{wire}\propto T^{-1/3}$ \cite{AAK82} and 
 $L_{\Phi}^\mathrm{ring}\propto T^{-1/2}$~\cite{LM04,TexMon05b,TexDelMon09}.

\begin{figure}[h]
\includegraphics[width=8cm]{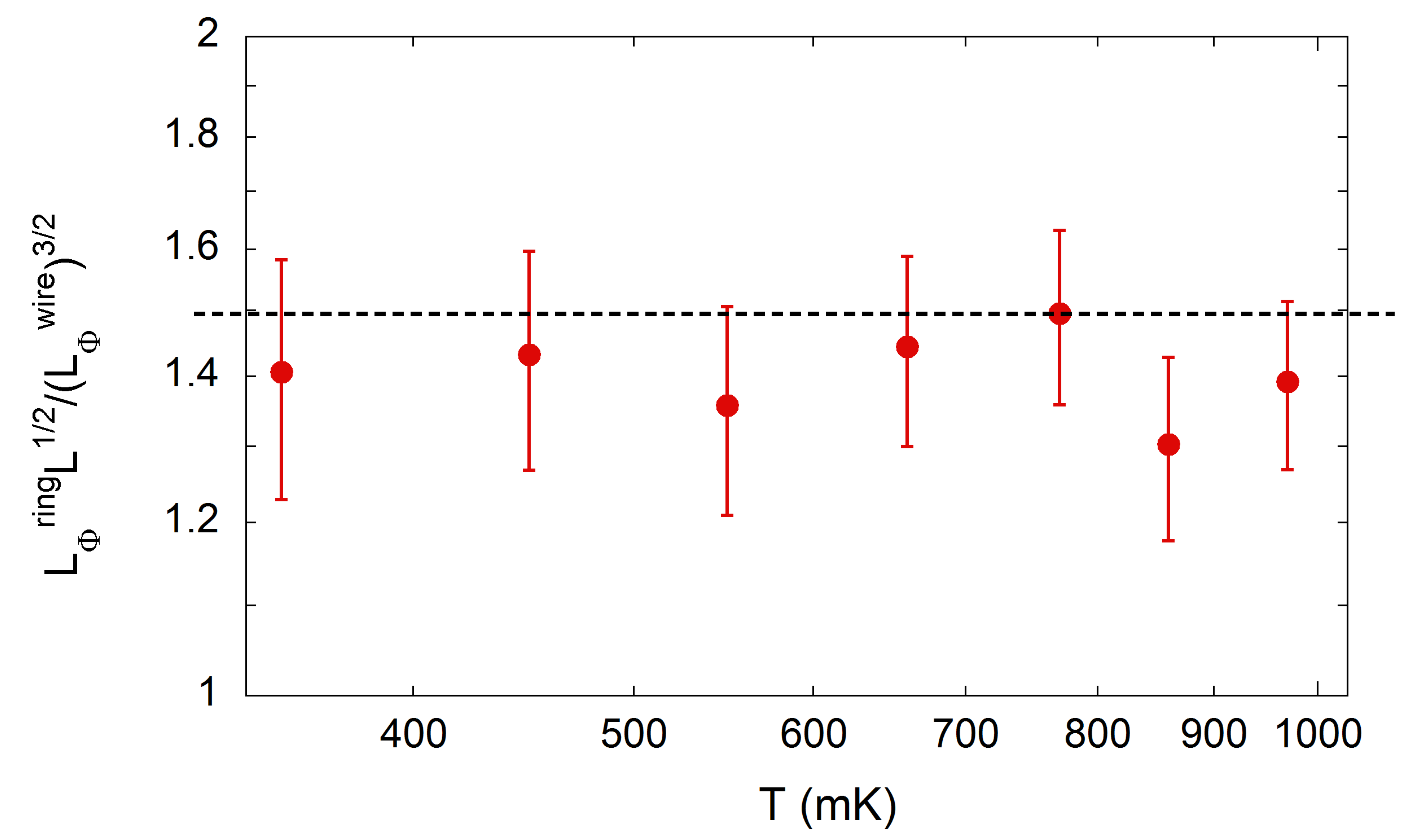}
\caption{
Experimental test of Eq.(\ref{eq:RelationLwireLring}), from the phase coherence lengths extracted from AB oscillations of the ring and WL of the wire.
The dotted line correspond to $2^{9/4}/\pi\simeq1.514$.
}
\label{fig:RelationLengths}
\end{figure}

We now recall simple arguments to understand the different exponents~\cite{LM04,TexMon05b,FerRowGueBouTexMon08,TexDelMon09}.
In Ref.~\cite{AAK82}, it is argued that decoherence results from the randomization of the phase of a given electron by the fluctuating electric field created by other electrons.
Within this picture an electron receives a phase $\Phi(t)\sim\int_0^t\D\tau\,V(\tau)$ from the electric potential whose fluctuations are characterized by the fluctuation-dissipation theorem $\int\D\tau\smean{V(\tau)V(0)}\sim e^2T\mathcal{R}_t$, where $\mathcal{R}_t$ is the resistance of the part of the sample probed by the electron during a time scale $t$. It can be written $\mathcal{R}_t\sim{}x(t)/s\sigma_0$, where $s$ is the section of the wire, $\sigma_0$ the Drude conductivity and $x(t)$ the distance covered by the electron.
The electronic phase presents a behaviour with time given by
$\smean{\Phi(t)^2}\sim\int_0^t\D\tau\int_0^t\D\tau'\,\smean{V(\tau)V(\tau')}\sim e^2T\,t\,x(t)/s\sigma_0$.
In a long wire, the motion is of diffusive nature, $x(t)\sim\sqrt{Dt}$ leading to the phase diffusion 
$\smean{\Phi(t)^2}\sim e^2T\,t^{3/2}\,\sqrt{D}/s\sigma_0$. We extract the relevant time scale by writing $\smean{\Phi(t)^2}\sim(t/\tau_\Phi^\mathrm{wire})^{3/2}$ with $\tau_\Phi^\mathrm{wire}\propto T^{-2/3}$.
In a ring, the winding trajectories are ergodic and the length $x(t)$ is simply the size of the system; we thus obtain the behaviour 
$\smean{\Phi(t)^2}\sim e^2T\,t\,L/s\sigma_0\sim t/\tau_\Phi^\mathrm{ring}$, leading to the time scale $\tau_\Phi^\mathrm{ring}\propto T^{-1}$.
We may write the relation between the two times as 
$\tau_\Phi^\mathrm{ring}\sim(\tau_\Phi^\mathrm{wire})^{3/2}/(\tau_D)^{1/2}$, where $\tau_D=L^2/D$ is the Thouless time. 
The precise dimensionless factor has been obtained by a careful analysis of the MC curve in 
Refs.~\cite{TexMon05b,FerRowGueBouTexMon08,TexDelMon09,Tex07b}: 
\begin{eqnarray}
   \label{eq:RelationLwireLring}
    L_{\Phi}^\mathrm{ring}=\frac{2^{9/4}}{\pi}\frac{(L_{\Phi}^\mathrm{wire})^{3/2}}{L^{1/2}}  
\end{eqnarray}
where $L_\Phi=\sqrt{D\tau_\Phi}$.
On Fig.~\ref{fig:RelationLengths} we check that $L_{\Phi}^\mathrm{ring}(T)$ and $L_{\Phi}^\mathrm{wire}(T)$ extracted from the experiment obey relation~(\ref{eq:RelationLwireLring}) with a very good accuracy.
The experimental verification of this relation definitely proves that the two temperature dependences of $L_\Phi(T)$ for the wire and for the ring emerges from the same mechanism described in a coherent picture~\cite{LM04,TexMon05b,TexDelMon09}.

Finally we comment on an issue which has been debated in the past~\cite{AB02,TexMon05b}~: for a given geometry, are the phase coherence length obtained from weak localization (WL) and conductance fluctuations (CF) identical~? 
In Ref.~\cite{FerRowGueBouTexMon08}, the measurement of the WL of a large array of rings has led to $L_{\Phi}^\mathrm{ring,WL}\propto T^{-1/2}$. This corresponds {\it quantitatively} to $L_{\Phi}^\mathrm{ring,CF}\propto T^{-1/2}$ obtained here.
These two measurements therefore give the first experimental demonstration that the phase coherence lengths extracted from WL and CF are indeed the same~\cite{footnoteSchopfer07}.

In conclusion, we have measured the phase coherence length $L_{\Phi}$ in samples etched in 2DEG of two different geometries by analyzing quantum corrections to the magnetoconductance. 
We have been able to extract for the first time the temperature dependence of the phase coherence length from the AB harmonics of a single diffusive ring.
The phase coherence length obtained from this method presents the behaviour $L_{\Phi}^\mathrm{ring}\propto T^{-1/2}$
different from the one obtained from the MC of a long wire,
$L_{\Phi}^\mathrm{wire}\propto T^{-1/3}$.
This demonstration of the geometrical sensitivity of the decoherence process by electronic interactions emphasizes the very precise understanding of the role of electronic interactions on low temperature phase coherent properties of metals.
In the present work, we have been able to investigate ergodic trajectories in a non ergodic temperature regime, $T>1/\tau_\Phi>1/\tau_D$, thanks to the selection of winding trajectories by the magnetic field. An experimental exploration of lower temperature is still desired: when $\tau_\Phi>\tau_D$, quantum interferences involve large scale properties which affect the decoherence process~\cite{TexDelMon09}. The investigation of the 0D regime~\cite{SivImrAro94,Treiber09,TreTexYevDelLer11}, $1/T>\tau_D$, still remains a challenging issue.

\acknowledgments
We acknowledge helpful discussions with H.~Bouchiat, M.~Ferrier,  T.~Ludwig, A.~D.~Mirlin, M.~Treiber,  J.~von Delft and O.~M.~Yevtushenko. This work has been supported by the European Commission FP6 NMP-3 project 505457-1 \textquotedblleft Ultra 1D\textquotedblright~and the \textsl{Agence Nationale de la
Recherche} under the grant ANR PNano \textquotedblleft QuSpin\textquotedblright.



\begin{thebibliography}{99}

\bibitem[*]{Mail}{Mail to: \texttt{saminadayar@grenoble.cnrs.fr}}

\bibitem{Imry}
Y. Imry, {\it Introduction to mesoscopic physics}, Oxford University Press (1997).

\bibitem{MontambauxBook}
E. Akkermans and G. Montambaux, {\it Mesoscopic physics of electrons and photons}, Cambridge University Press, Cambridge (2007).

\bibitem{ChaSch86}
S.~Chakravarty and A.~Schmid, Weak localization: the quasiclassical theory of
  electrons in a random potential, Phys. Rep. {\bf 140}, 193 (1986).

\bibitem{PieGouAntPotEstBir03}
 F.~Pierre, A.~B. Gougam, A.~Anthore, H.~Pothier, D.~Esteve and N.~O. Birge,
  Dephasing of electrons in mesoscopic metal wires, Phys. Rev.~B {\bf 68},
  085413 (2003).

\bibitem{Saminadayar07}
L. Saminadayar, P. Mohanty, R. A. Webb, P. Degiovanni and C. B{\"{a}}uerle, 
  Phase coherence in the presence of magnetic impurities,
  Physica E \textbf{40}, 12 (2007).

\bibitem{AAK82}
B. L. Altshuler, A. G. Aronov and D. E. Khmelnitsky, 
  Effects of electron-electron collisions with small energy transfers on quantum localisation, 
  J. Phys. C: Solid St. Phys. \textbf{15}, 7367 (1982).

\bibitem{LM04}
T. Ludwig and A. D. Mirlin, 
  Interaction-induced dephasing of Aharonov-Bohm oscillations,
  Phys. Rev. B \textbf{69}, 193306 (2004).

\bibitem{TexMon05b}
C. Texier and G. Montambaux, 
  Dephasing due to electron-electron interaction in a diffusive ring,
  Phys. Rev. B \textbf{72}, 115327 (2005).

\bibitem{TexDelMon09}
C. Texier, P. Delplace and G. Montambaux, 
  Quantum oscillations and decoherence due to electron-electron interaction in networks and hollow cylinders,
  Phys. Rev. B \textbf{80}, 205413 (2009).

\bibitem{Treiber09}
M. Treiber, O. M. Yevtushenko, F. Marquardt, J. von Delft and I. V. Lerner, 
Dimensional Crossover of the dephasing time in disordered mesoscopic rings,
  Phys. Rev. B \textbf{80}, 201305 (2009).

\bibitem{TreTexYevDelLer11}
M. Treiber, C. Texier, O. M.~Yevtushenko, J. {von~Delft} and I. V. Lerner,
Thermal noise and dephasing due to electron interactions in non-trivial geometries,
  Phys. Rev. B \textbf{84}, 054204 (2011).

\bibitem{AltAro81}
B.~L. Al'tshuler and A.~G. Aronov, 
  Magnetoresistance of thin films and of wires in a longitudinal magnetic field, 
  JETP Lett. {\bf 33}, 499 (1981).

\bibitem{RefAB}
The AB effects 
[Y. Aharonov and D. Bohm,
  Significance of electromagnetic potentials in the quantum mechanics,
  Phys. Rev. {\bf 115}, 485--491 (1959)]
was in fact proposed earlier in:
W. Ehrenberg and R. E. Siday,
  The refractive index in electron optics and the principles of dynamics,
  Proc. Phys. Soc. (London) B {\bf 62}, 8--21 (1949).

\bibitem{WasWeb86}
S.~Washburn and R.~A. Webb,
 Aharonov-Bohm effect in normal metal. Quantum coherence and transport,
 Adv. Phys. {\bf 35}, 375 (1986).

\bibitem{footnoteMeydi}
Different temperature dependent phase coherent lengths have been extracted in the recent analysis of the WL in samples made of large number ($10^6$) of rings arranged to form a grid~\cite{FerRowGueBouTexMon08}.

\bibitem{FerRowGueBouTexMon08}
M. Ferrier, A. C. H. Rowe, S. Gu\'eron, H. Bouchiat, C. Texier and G. Montambaux,  
  Geometrical dependence of the decoherence due to electronic interactions in a GaAs/GaAlAs square network, 
Phys. Rev. Lett. \textbf{100}, 146802 (2008).

\bibitem{YasuPRL09}
Y. Niimi, Y. Baines, T. Capron, D. Mailly, F.-Y. Lo, A. D. Wieck, T. Meunier, L. Saminadayar and C. B\"{a}uerle, 
 Effect of Disorder on the Quantum Coherence in Mesoscopic Wires,
  Phys. Rev. Lett. \textbf{102}, 226801 (2009)~;
Y. Niimi, Y. Baines, T. Capron, D. Mailly, F.-Y. Lo, A. D. Wieck, T. Meunier, L. Saminadayar and C. B\"{a}uerle, 
 Quantum coherence at low temperatures in mesoscopic systems: Effect of disorder,
  Phys. Rev. B \textbf{81}, 245306 (2010).

\bibitem{BeeHou88}
V.~K. Dugaev and D.~E. Khmel'nitzki{\u\i}, 
 Magnetoresistance of metal films with low impurity concentrations in parallel magnetic field, 
  Sov. Phys. JETP {\bf 59}, 1038 (1984)~;
C.~W.~J. Beenakker and H.~{Van~Houten},  
  Boundary scattering and weak localization of electrons in a magnetic field, 
  Phys. Rev.~B {\bf 38}, 3232 (1988)~;
H. Van Houten {\it et al}, 
 Boundary scattering modified one-dimensional weak localization in submicron GaAs/GaAlAs heterostructures,
  Surf. Sci. \textbf{196}, 144 (1988).

\bibitem{footnoteZNA}
In Ref.~\cite{NarZalAle02}, the study of the crossover between diffusive and ballistic regimes in 2D samples has shown that the decoherence rate is given by the addition of a term $\propto T$ (2D-diffusive) and a term $\propto T^2$ (ballistic) \cite{AAK82,FA83}.
We assume that diffusive and ballistic rates can also be added in the quasi-1D geometry.

\bibitem{NarZalAle02}
B. N. Narozhny, G. Zala and I. L. Aleiner,
  Interaction corrections at intermediate temperatures: Dephasing time,
  Phys. Rev.~B {\bf 65}, 180202 (2002).

\bibitem{FA83}
H. Fukuyama and E. Abrahams, 
 Inelastic scattering time in two-dimensional disordered metals,
  Phys. Rev. B \textbf{27} 5976 (1983).

\bibitem{Tex07b}
C. Texier,  
 Effect of connecting wires on the decoherence due to electron-electron interaction in a metallic ring, 
  Phys. Rev.~B {\bf 76}, 153312 (2007).

\bibitem{contacts}
V. Chandrasekhar, P. Santhanam and D. E. Prober, 
 Weak localization and conductance fluctuations in complex  mesoscopic geometries
  Phys. Rev. B \textbf{44}, 11203 (1991) ;
K. Kobayashi, H. Aikawa, S. Katsumoto and Y. Iye, 
 Probe-Configuration-Dependent Decoherence in an Aharonov-Bohm Ring,
  J. Phys. Soc.  Jpn. \textbf{71}, 2094 (2002).

\bibitem{footnoteLionel}
Measurements consistent with  $L_\Phi\propto T^{-1/2}$ in similar samples were mentioned in:
L.~Angers, E.~Zakka-Bajjani, R.~Deblock, S.~Gu\'{e}ron, H.~Bouchiat, A.~Cavanna, U.~Gennser and M.~Polianski, 
  Magnetic-field asymmetry of mesoscopic dc rectification in Aharonov-Bohm rings, 
  Phys. Rev.~B {\bf75}, 115309 (2007).

\bibitem{HanKriPedSorLin01}
Note that $L_\Phi\propto T^{-1}$ was extracted from AB oscillations of \textit{ballistic} rings in:
A. E. Hansen, 
A. Kristensen, S. Pedersen, C. B. S{\o}rensen and P. E. Lindelof,
  Mesoscopic decoherence in Aharonov-Bohm rings,
  Phys. Rev. B \textbf{64}, 045327 (2001).

\bibitem{AB02}
Ya. M. Blanter,
 Electron-electron scattering rate in disordered mesoscopic systems,
 Phys. Rev.~B {\bf 54}, 12807 (1996)~;
I. L. Aleiner and Ya. M. Blanter, 
 Inelastic scattering time for conductance fluctuations,
  Phys. Rev. B \textbf{65}, 115317 (2002).

\bibitem{footnoteSchopfer07}
The relation between AB oscillations (CF) and Altshuler-Aronov-Spivak oscillations (WL) \cite{AAS81} has been verified as a function of the sample size at a {\it fixed} temperature in \cite{Schopfer07}.

\bibitem{AAS81}
B. L. Al'tshuler, A. G. Aronov and B. A. Spivak, 
 The Aaronov-Bohm Effect in disordered conductors,
  JETP Lett. \textbf{34}, 94 (1981).

\bibitem{Schopfer07}
F. Schopfer, F. Mallet, D. Mailly, C. Texier, G. Montambaux, C. B\"auerle and L. Saminadayar, 
  Dimensional  crossover in quantum networks: from mesoscopic to 
         macroscopic physics,
  Phys. Rev. Lett. \textbf{98}, 026807 (2007).

\bibitem{SivImrAro94}
U.~Sivan, Y.~Imry and A.~G. Aronov, 
  Quasiparticle lifetime in a quantum dot,
  Europhys. Lett. {\bf 28}, 115--120 (1994).

\end{thebibliography}
\end{document}